\documentstyle[preprint,eqsecnum,aps]{revtex}
\tightenlines
\newcommand{\be}{\begin{equation}}
\newcommand{\ee}{\end{equation}}
\newcommand{\bea}{\begin{eqnarray}}
\newcommand{\eea}{\end{eqnarray}}

\begin{document}
\draft
\title{Barrier effects on the collective excitations of split
Bose-Einstein condensates}

\author{P. Capuzzi and E. S. Hern\'andez}

\address{Departamento de F\'{\i}sica, Facultad de Ciencias
Exactas y Naturales, \\
Universidad de Buenos Aires, RA-1428 Buenos Aires, \\
$^1$ and Consejo Nacional de Investigaciones Cient\'{\i}ficas y
T\'ecnicas, Argentina}

\maketitle

\begin{abstract}
 We  investigate the  collective excitations of a single-species Bose
gas at $T=0$ in a harmonic trap where the confinement  undergoes some
splitting along one spatial direction. We mostly consider onedimensional
potentials consisting of two harmonic wells separated a distance $2 z_0$, since they  essentially contain all the barrier effects that one may visualize  in the 3D situation.
We find, within a hydrodynamic approximation, that regardless the
dimensionality of the system, pairs of levels in  the excitation
spectrum, corresponding to neighbouring even and odd excitations, merge together as one 
increases the  barrier height up to the current value of the chemical potential.  The excitation spectra computed in the hydrodynamical or Thomas-Fermi limit are compared with the results of exactly solving the time-dependent Gross-Pitaevskii equation. We analyze as well the characteristics of the spatial pattern of  
excitations of threedimensional boson systems according to the amount of   splitting of the condensate.

\end{abstract}
\vspace{0.5truecm}
Pacs{\,\,03.75.Fi,05.30.Jp,32.80.Pj}
\narrowtext

\section{Introduction}

The experimental realization of Bose-Einstein condensation  (BEC)\cite{Rb,Na,Li}
has inspired an impressive amount of further laboratory research together with important  theoretical advances. The overwhelming material produced in the last
three years has been recently reviewed by several authors \cite{PW98,Shi,Dalfo98}.
Among the various fascinating subjects that have attracted the attention of the 
community,  the predictions of the spectrum of low-lying excitations and sound 
velocity\cite{ST1,ST2} in a hydrodynamic picture based on perturbations over the Thomas-Fermi regime, and their experimental verification\cite{Col1,Col2,Col3,sound} 
play a special role, since the excellent agreement between theoretical and observed  figures bring into evidence the validity of the mean field description  of large condensates. Methods to obtain exact solutions of the time-dependent 
Gross-Pitaevskii equation (GPE) have been developed\cite{metodos} which permit 
further confirmation of  the accuracy  of the TF+hydrodynamic  approach in the 
thermodynamic limit; moreover, the large amplitude motion of a condensate has 
been investigated in a fashion that demonstrates the validity of a scaling 
hypothesis\cite{scaling}, according to which a condensate may evolve mantaining 
the appearance of a barometric distribution in a quadratic field. 

A particular subject of the BEC field is the possibility of forming and 
studying double condensates from several viewpoints. Starting from the 
original MIT trap \cite{Na} and its later improvements that allowed, for 
instance, the observation of interference fringes and the experimental 
realization of the coherence of matter waves\cite{interferencia}, the 
isomorfism between atomic condensates and optical devices has been 
intensively investigated\cite{GMZM}.  Tunneling between the wells and 
the features of Josephson currents now constitute common subjects in 
the literature\cite{tuneleo,Josephson}.  A special chapter includes the 
binary mixtures of condensates or double condensates of distinguishable 
particles, typically associated to hyperfine splitting of the magnetic 
atom levels in the  trap\cite{doble}. In particular, a mean field description of two species, spatially split double condensates has been advanced in Ref. \cite{HF}.     Indeed, most theoretical approaches 
regarding the low energy dynamics of a double well\cite{GM97,GW98},   
tunneling and/or Josephson effects consider essentially this type of 
distinguishable systems - or concentrate upon  dynamical aspects 
that exclude the quantum and thermodynamic analysis of the double well 
as a subject in its own right.  The investigation of the properties of 
single-species, double condensates with substantial spatial  overlapping 
could, however, become relevant in the near future in view of the current 
possibility of adiabatically distorting traps\cite{adia} creating long-lived 
systems with various minima and barriers \cite{lattice}.  

	Recently, we have advanced some results on the thermodynamics of 
BEC of indistiguishable particles in harmonic double wells\cite{Pablo}, which indicate that with a 
minimum amount of complication with respect to the well-known harmonic 
traps, one can trace the evolution of the atomic spectrum as a single 
potential becomes progressively split into two with the same frequency, 
at a distance 2$z_0$ apart along the $z$-axis, and separated by a barrier 
$V_0=M\,\omega_z^2 \,z_0^2/2$. The major features of BEC, under such 
conditions,  can be accurately described and interpreted in terms of a 
smooth transition from a system with $N$ particles to two identical 
ones with $N/2$ each. With this in mind, in the present work we intend 
to analyze the effects of the presence of such barriers upon the collective 
excitations of the condensate. For this sake, in Sec. 2 we concentrate 
on the onedimensional, harmonic double well problem; we extract the 
exact eigenspectrum of the linearized hydrodynamical equation\cite{ST1}, 
and confront the results thus obtained with the exact Fourier spectrum 
of the GPE solutions. In Sec. 3, we examine the  threedimensional system 
and examine, in particular, the spatial patterns of the perturbation as 
the well splitting increases. Section 4 is devoted to the full discussion 
of the results here presented. The final summary and comments are the 
subject of Sec. 5.

\section{Onedimensional Bose system in a harmonic double well}

The general formalism leading from the time dependent Gross-Pitaevskii
equation (GPE)  to the description of the low energy excitations of a
Bose-Einstein condensate has been derived by Stringari \cite{ST1} and
reviewed, for example, in Refs. \cite{PW98,Dalfo98}. The procedure consists
of linearizing the hydrodynamical version of the GPE with respects to small
deviations $\delta \rho$ from the Thomas-Fermi equilibrium density
inside the condensate
\begin{equation}
\rho_{TF}({\bf r})= \frac{\mu -V_{ext}({\bf r})}{g}
\end{equation}
where $\mu$ is the chemical potential evaluated in the Thomas-Fermi regime
and $g= 4 \pi \hbar^2\,a/M$ is the hard-sphere interaction strength
for an $s$-wave scattering length $a$. This linearized equation reads
\be
M\,\Omega^2\,\delta\rho = -\nabla . \left[(\mu-V_{ext})\,\nabla\,\delta\rho\right]
\label{hidro3D}
\ee

 Its one dimensional version, namely  
\be
M\,\Omega^2\,\delta\rho = -\frac{\displaystyle \partial}{\displaystyle \partial
 z}\left[(\mu-V_{ext}(z))\,\frac{\displaystyle\partial }{\displaystyle \partial z}\,
\delta \rho\right]
\label{hidro1D}
\ee
with $V_{ext}$ a harmonic potential with frequency $\omega$, after an adequate
scaling is precisely the differential equation satisfied by the Legendre
polynomials $P_n(\sqrt{M \omega^2/2 \mu_0}\,z)$, the corresponding eigenvalues
being of the form $\Omega_n^2 = \omega^2 n (n+1) /2$\cite{exacta}. Furthermore, the TF chemical potential $\mu_0$ reads, in this case,
\be
\mu_0=\frac{\hbar\,\omega}{2}\,\left(6 \pi\,N\,\frac{a}{a_z}\right)^{2/3}
\ee
with $a_z=\sqrt{\hbar/M\,\omega}$ the corresponding oscillator length.

Our current purpose is to find the exact solutions of Eq. (\ref{hidro1D})
when $V_{ext}$ is the double well potential
\be
V_{ext}=\frac12\,M\,\omega^2(|z|-z_0)^2
\label{pote1D}
\ee
In this case, within the TF approximation, the ratio of the chemical potential $\mu$ to that of a single
well is found to satisfy the relationships
\begin{eqnarray}
&&\frac{\mu}{\mu_0}=
\frac{\displaystyle 1}{\displaystyle 2^{2/3}} \,\,\,\,{\rm if} \;\;\; V_0 \geq \mu
\nonumber
 \\
\label{mu1D}
\\
&&\left(\frac{\mu}{\mu_0}\right)^{3/2}+\frac34\,\left(\frac{V_0}{\mu_0}
\right)^{1/2}\,\left(\frac{\mu}{\mu_0}\right)-\frac12\,
\left(\frac{V_0}{\mu_0}\right)^{3/2}= 1 \,\,\,\, {\rm otherwise}
\nonumber
\end{eqnarray}
where   the barrier  effect  is fully  included in  the ratio $V_0/\mu_0$. The decrease $\mu/\mu_0$ of the chemical 
potential is plotted as a function of the dimensionless variable 
$\sqrt{V_0/\mu_0}=z_0/Z_0$, with $Z_0$ the single-condensate size, in Fig. 1, 
 while in Fig. 2 we display   three TF densities,
respectively  corresponding to $\mu$ = 5/2 $V_0, V_0$ and 1/4 $V_0$.

We now look for the solutions of Eq. (\ref{hidro1D}) in a fashion  that
closely parallels the computation of the eigenvalues of
Schr{\"o}dinger equation for the double well hamiltonian \cite{merzbacher}.
In other words, we look for eigenvalues $\Omega_{\nu}^2$ and eigenvectors
$\delta \rho_{\nu}$ of the form
\be
\Omega_{\nu}^2 = \frac{\omega^2}{2} \nu\,(\nu+1)
\label{Omega}
\ee
and
\be
\delta \rho_{\nu} = P_{\nu}\left[\sqrt{M \omega^2/2 \mu}\,(|z| -z_0)\right]
\label{Pnu}
\ee
with $P_{\nu}(x)$ a generalized Legendre function \cite{Grad}. To obtain the
complete solution of the hydrodynamic problem  we observe that i) the operator on the right-hand side of (\ref{hidro1D}) has well defined parity and ii)  the boundary condition that guarantees real
values of $\Omega$ is the nondivergence of the density fluctuation  on the
edges of the condensate \cite{FliesserA}. Taking this into account, we construct
the eigensolutions (\ref{Pnu}) separately for positive and negative $z$ and
request the matching conditions at $z=0$\cite{merzbacher},
\be
 P_{\nu}\left[-\sqrt{M \omega^2/2 \mu}\, z_0\right]=0\,\,\,\,{\rm for \,\,odd \,\,solutions},
\ee
\be
P'_{\nu}\left[-\sqrt{M \omega^2/2 \mu}\,z_0\right]=0\,\,\,\,{\rm for \,\,even \,\,solutions}.\ee

When the barrier height reaches the current chemical potential value $\mu$,
Eq. (\ref{hidro1D}) is well defined in two separate domains (cf. Fig. 2)  and we can
construct the density fluctuations as even and odd superpositions of the
exact single-well solutions.
\bea
\delta\rho_{\rm even}&=&{\cal
N}\,\left\{\delta\rho_0(z-z_0)\,\Theta(z)+
\delta\rho_0(-z-z_0)\,\Theta(-z)\right\} \nonumber \\
\label{merge} \\
\delta\rho_{\rm odd}&=&{\cal N}\,
\left\{\delta\rho_0(z-z_0)\,\Theta(z)-
\delta\rho_0(-z-z_0)\,\Theta(-z)\right\}
\nonumber
\eea
being $\delta\rho_0$ one eigensolution for the 1D harmonic trap 
with a length scale corresponding to halving the particle number, and
$\Theta(x)$ the Heavyside function. Thus, once the condensate is split
into two pieces, under the TF approximation, the excitation spectrum for
any finite barrier $V_0$ higher than the ground state energy $\mu$  is
the single well one, however  doubly degenerate. The appearance of the
1D excitation spectrum as a function of the separation
$\sqrt{V_0/\mu}=z_0/Z$
is depicted in Fig. 3; moreover, to inspect the departures in the behavior of the excitations from the
hydrodynamic limit, we have integrated  the time dependent GP equation and 
display with symbols some indicative results.
This integration is achieved in two steps: we first build up
the condensate wavefunction solving the time-independent problem 
by the steepest descent method \cite{Dalfo96}, and afterwards we add some 
noise to the condensate density $\rho_0$.  We then follow the time evolution
of the perturbed density and compute the fast Fourier transform (fft) at  one  minimum of the double well potential,  which provides  the spectrum of  
small amplitude oscillations of the double condensate \cite{Col4}. This is illustrated
in Fig. 4, where we plot this spectrum for four choices of the ratio
$\sqrt{V_0/\mu}$ and an interaction strength  $N\,a/a_z=10$. We realize that for very low barriers, the spectrum is the one 
corresponding  to a single well, but as one increases the separation 
between the minima,  two quasidegenerate branches can be distinguished. These branches   completely merge at infinitely large barriers, however, they remain observable at  finite 
values of $V_0$ due to the limited resolution on the fft calculation. We notice that although the slopes of the odd eigenvalues of the TF hydrodynamics exhibit a sharp discontinuity at the barrier limit $V_0/\mu=1$, the fft results behave smoothly across the transition. We shall analyze this issue later in Sec. 4.

\section{The threedimensional Bose system}

We now discuss the solution of the full equation (\ref{hidro3D}) for a
cylindrical potential of the form
\be
V_{ext}({\bf r}) = \frac{M\,\omega_r^2}{2}\,r^2+\frac{M\,\omega_z^2}{2}
\,(|z|-z_0)^2
\ee
with a given aspect ratio $\lambda=\omega_z/\omega_r$. In this case, the
chemical potential $\mu$ in the Thomas-Fermi limit is related to the
single-well one given by
\be
\mu_0=\frac{\hbar\,\tilde{\omega}}{2}\,\left(15\,N\,\frac{a}{\tilde{a}}\right)^{2/5}
\ee
with $\tilde{\omega} = (\omega_r^2\,\omega_z)^{\frac{1}{3}}$ and $\tilde{a}$ the corresponding oscillator length, according to
\begin{eqnarray}
&&\frac{\mu}{\mu_0}=
\frac{\displaystyle 1}{\displaystyle 2^{2/5}} \,\,\,\,{\rm if} \;\;\; V_0 > \mu 
\nonumber
\\
\label{mu3D}
\\
&&\left(\frac{\mu}{\mu_0}\right)^{5/2}+\frac{15}{8}\,\left(\frac{V_0}{\mu_0}\right)^{1/2}\,\left(\frac{\mu}{\mu_0}\right)^{2}
-\frac{5}{4}\,\left(\frac{V_0}{\mu_0}\right)^{3/2}\,\left(\frac{\mu}{\mu_0}\right)+\frac{3}{8}\,
\left(\frac{V_0}{\mu_0}\right)^{5/2}=1\,\,\,\, {\rm otherwise}
\nonumber
\end{eqnarray}
The ratio $\mu/\mu_0$ is displayed in Fig. 5 as a function of the
dimensionless variable $\sqrt{V_0/\mu_0}$ and in Fig. 6, we display three contour plots of a TF density - {\it i.e.}, the condensate edge and two 
inner equiprobability contours - for a number of particles such that 
$\mu= 5/2\,V_0$. Alternatively, these curves represent  three different condensate edges for $\mu= 5/2\,V_0, V_0$ and $1/4\,V_0$, in correspondence with the 1D density profiles of Fig. 2.

Due to the different definition of the $z-$components of the restoring force
for positive and negative values of the coordinate, standard methods
\cite{ST1,ST2,FliesserA,Csordas} employed to derive the eigenfunctions and eigenvalues
for anisotropic single-well potentials are not especially useful in this
case. We have rather found convenient to expand the solutions $\delta \rho(r,z)$
in the  spherical basis $P^{(2n)}(r)\,r^l\,Y_{lm}(\theta, \phi))$ \cite{ST1},
and diagonalize the matrix representing  the operator $\nabla . \left
[(V_{ext}-\mu)\,\nabla \right]/M$. The eigenspectra thus obtained are labelled
by the magnetic quantum number $m$ and in order to illustrate the appearance
of the spectra $\Omega_{mk}$, with $k$ just numbering the level for the
given $m$, in Fig. 7 we display these energies for (a)  $m=0$ and (b)
 $m=1$ as functions of the well separation. We appreciate the general 
trend similar to that already analyzed in the 1D case, namely, the gradual 
merge of pairs of levels corresponding to even and odd eigenfunctions
and their collapse into the doubly degenerate spectrum of single-well excitations. The apparent discontinuity of the odd level energies when the
chemical potential equals the barrier height is also visible; furthermore,
the eigenfrequencies arising from the numerical diagonalization can be
compared with those provided by the Fourier transformation of an initial
perturbation that evolves according to the full GPE dynamics.  Figure  8
displays a typical Fourier spectrum for a weak barrier with $\sqrt{V_0/\mu}=0.14$; some
eigenvalues of the dynamical matrix are indicated on the abscissa axis so
as to remark the good agreement between both procedures, in spite
of the inaccuracies intrinsic to the Thomas-Fermi approximation.
However, the way in which the excitation was created in the fft method - {\i.e}, adding some random fluctuation to the TF ground state - 
does not allow us to control exactly the particular selection  of the fluctuating modes or their 
relative intensity. Thus, there exist eigenfrequencies of the
problem that we cannot clearly see in the fft spectrum of an arbitrary initial condition like the typical example in Fig. 8.

   The shape of the perturbation $\delta \rho_{mk}(r,z)$ as a function
of the coordinates inside the condensate deserves some comments. We have
chosen to analyze the transition density patterns in the following way:
keeping Fig. 6 as a reference, we consider, for a given condensate characterized by a choice of the chemical potential $\mu$, the equiprobability contours
$\rho_{TF}(r,z)=\rho_0$  with $\rho_0$ a constant that vanishes at the condensate edge and increases towards the center of the trap. In Fig. 9, we are drawing 
four portraits corresponding to the first odd parity excitation for $m$=0, for different well separations (or to different chemical 
potentials for fixed barrier height). Each of these  sets contains  curves 
$\delta \rho[r(z),z]$ as functions of $z$ for a selected equiprobability contour
$\rho_0$.
Full lines indicate the condensate edge $\rho_0$= 0 , while the
different dashed lines correspond to inner contours of increasing equal density. Figures 10  to 12 respectively display
the same patterns for the first even parity excitation with $m$=0 and
for the first odd and first even parity eigenmodes with $m$=1.   Very similar scenarios are obtained for other low energy  levels, the present choice being convenient to observe the
evolution of  the perturbation and its nodal characteristics in terms of
relatively simple, however rich enough, spatial structure. We examine these results further in Sec. 4.

\section{Discussion of the results}

From the  previous chapters, we recognize that the most important differences
between the present collective dynamics and the well-known excitation spectra
of Bose condensates in anisotropic harmonic wells, can be fully understood in
terms of the dynamical properties of the 1D double well. It is interesting to
remark that, in this situation,  the hydrodynamical wave equation (\ref{hidro1D})
can be solved almost analytically in the same spirit, and with the same degree of
complexity, as the quantum problem of the Schr{\"o}dinger eigenspectrum of a
harmonic double well.
The behavior of the excitation energies under a progressive well splitting
(cf. Figs. 3 and 6) is a monotonic decrease together with an evolution
towards doubly degeneracy of the excitations of a single harmonic well.  This  can be understood in terms  of a simple classical picture that 
illustrates the effect of the barrier; in fact, the hydrodynamic equation (\ref{hidro1D})
may be interpreted as the classical wave equation  of a perturbing
density fluctuation  $\delta\rho$ travelling inside the condensate with
a local velocity  $v(z)=\sqrt{\left(\mu-V_{ext}(z)\right)/M}$. With this in mind, we  can estimate the time $\tau$ employed by the perturbation
to go from one well to the other  and the oscillation period $T$, namely,
\bea
\tau&=&\int_{-a}^{a}\,dz\,\frac{\displaystyle 1}{\displaystyle
v(z)}=\frac{\displaystyle 2\,\sqrt{2}}{\omega}\,{\rm Arcsin}\left(\frac{V_0}{\mu}\right)\\
T&=&2\,\int_{-a-Z}^{a+Z}\,dz\,\frac{\displaystyle 1}
{\displaystyle v(z)}=\frac{2 \pi \sqrt{2}}{\omega}\left[
1+\frac{\displaystyle 2}{\displaystyle \pi}\,{\rm Arcsin}\left(\frac{V_0}{\mu}\right)\right]
\nonumber
\\
&=& T_0+2\,\tau 
\label{Tclasi}
\eea 
being $T_0$ the period of the orbit of energy $\mu$ for a particle with
mass $2 M$  in a single well of frequency $\omega$; one may notice that under the replacement $p=M\,\omega\,r$, Fig. 6 describes  as well the classical phase portrait of a particle in the 1D-double well.  We realize then that
when $V_0$ is sufficiently low, the fluctuation essentially oscillates
around the condensate along the classical orbit with period $T_0$; in
particular, when $V_0$ equals $\mu$, the excitation requires twice a
period to cross the double condensate, the classical orbit -pictures in medium dashes in Fig. 6 - being the
separatrix between the two different oscillation regimes. As $V_0$ grows up,
a delay equal to 2 $\tau$ appears in the oscillation time, reflecting the 
finite amount of time requested to switch back and forth between the two wells.

The decrease of the eigenfrequencies  can thus be unambiguously interpreted 
  in terms of the enlarging of the classical period of the orbits due to the presence of a finite barrier. However, the most striking outcome of the TF computation is the discontinuity of the odd-state engenfrequencies as we approach the limiting regime $\mu-V_0$ from above. In fact, contrasting
this behavior with the strict Fourier spectrum, for which some results are
indicated with symbols  in Fig. 3, we realize that this discontinuity is an
artifact of the TF approach, removable by an exact calculation. The
explanation is as follows: let us project  the wave equation
onto an eigenstate, thus expressing the eigenvalues in the form
\be
\Omega_{\nu}^2=\frac1M\,\int d{z}\,\left\vert \frac{\displaystyle d \delta
\rho_{\nu}}{\displaystyle d z}\right\vert^2\,\left[\mu- V_{ext}({\bf r})\right] 	
\ee
where the integration is to be performed within the borders of the TF
condensate. If $\mu$ is lower than the barrier $V_0$, the condensate is
split into two disjoint condensates and the region around $z=0$ is excluded
from the integration domain. This exclusion is not significant for even
parity states, since the density gradient vanishes at $z=0$; however,
this is not the case for   the odd parity levels, whose exact Fourier spectrum,
lying higher than the TF one (cf. Fig. 3) includes the contributions to
the eigenfrequencies which stem from the whole spatial domain, in particular,
the subbarrier region where the exact Gross-Pitaevskii wave function yields
finite  probability densities for the trapped bosons.

This issue can  be also easily interpreted taking into account that  if one excites
the condensate, generating a small density fluctuation with even parity, 
this excitation will oscillate in time within the limits of each
well, basically unaffected by the presence of the other.
By contrast, if the fluctuation is an odd parity one, a peak in one 
side of the condensate is imaged by a dip in the other; subsequent time  evolution  forces this pattern  to oscillate from 
one side to the other, crossing the barrier. Now,  since within the 
TF approximation both the condensate and its fluctuations are bounded to the
region  $V_{ext} < \mu$, we recognize that  when the barrier increases above
the chemical potential value, any pertubation is forced to remain on its side with no posible interwell transition - thus  the levels correspond to  two 
independent Bose-Einstein condensates-. This is not what happens in 
a real double well described by the full GPE, where  both  condensates interpenetrate.

The 3D situation gives us the opportunity to investigate  transition density
profiles for different selections of labels and for several possible manners
of disecting the condensate. In Sec. 3 we have displayed  some of the spatial
characteristics of the first odd and even $m$=0 and 1 modes (in Figs. 9 to 12),
 choosing the variable $z$ to parametrize various equiprobability
contours inside the condensate. The evolution of these patterns illustrates
several interesting facts. We first realize that the spatial distribution of
the perturbation evolves, with increasing well separation,  so as to eliminate
nodes of the eigenfunction, in order to account for the level merging visible
in Figs. 3 and 7.   Secondly, we observe the transformation from a 
pattern localized in one harmonic potential to a  pattern of the
same nature, however split into the two disjoint wells. In particular,  the fourth portrait of each of these figures corresponds to an almost split condensate.  We notice that  in neither case the exact form of the twofold distribution after separation coincides with the original 
one, due to the fact that in the large $V_0$ limit, the fluctuation is 
of the form (\ref{merge}) with the parity of $\delta \rho_0$ 
uncorrelated to  that of the starting $\delta \rho$. Thus, should one duplicate the original pattern in each well after their definitive
splitting,   the energy would be  higher. We observe that the $m$=0 modes 
display finite density fluctuations in the outer edges of the condensate 
($z=\pm(z_0+Z)$) whereas in $z=0$, where the inner boundary surface grows  
up with increasing well separation, the density vanishes or not  according to 
the parity being with odd
or even;  by contrast, in either case the $m$=1 modes intend to concentrate 
the fluctuations at values of $z$ far from either the outer ({\it i.e.},
$z \approx \pm(z_0+Z))$ or the inner   ($z \approx 0$) condensate  edges.  The effect of the barrier upon even parity perturbations is to introduce a discontinuity in the slope of $\delta \rho$, similar to that in the TF density itself.  We 
also see that, as in the single-well problem, the inner contours, where 
the above  density is larger, experience weaker density fluctuations  
than  the condensate edge.

\section{Summary}

In this work we have discussed  the characteristics of
the small amplitude collective excitations of a Bose condensed gas  in 
harmonic wells where the confinement has undergone some splitting along 
one of the three directions. The  excitations of the confined bosons has been investigated  within the hydrodynamic limit of the GPE (TF limit), in other words, we have   analyzed the 
small fluctuations of the condensate density in the limit of large
number of particles. This enables us to 
determine the chemical potential as a function of the barrier height and
the single-well parameters in the 1D and 3D problems. We have considered extensively the 1D potential consisting of two harmonic wells separated a distance 2$a$, since it contains essentially all the barrier effects that one may visualize later in the 3D situation.

  The general results can be summarized as follows. We
find that the energy levels corresponding to the density fluctuations gradually merge into pairs as one increases the barrier height;  within the TF approximation, as soon as the barrier  overcomes the current value of the chemical potential the spectrum becomes completely degenerate.  We also have shown that the behavior of the eigenspectrum as a function of the well separation  can be interpreted, in the 1D case,  by means of a classical picture,  taking into consideration the way in which the finite barrier affects the period of the  classical orbit.
The appearance of degeneracy in the eigenspectrum, that images the well-known property of the Schr{\"o}dinger eigenspectrum of a 1D harmonic double potential, gives rise to various consequences upon the  spatial pattern
 of the excitation near the surface of a 3D condensate when the latter has 
undergone splitting.   In the 3D problem we have displayed typical excitations 
corresponding to levels with and without z-proyection of the angular momentum 
and have discussed the consequences of progressively splitting the condensate.
We recognize the usual differences between  fluctuations with zero and
nonzero magnetic quantum number; while the first ones remain finite at the edges
$z=\pm (z_0+Z)$ (cf. Figs. 9 and 10),  they vanish at those positions if
$m \neq $ 0. These features remain valid as one increases the barrier height 
and splits the condensate into two pieces.

 To summarize, we remark that the collective excitations  of this bistable  system  can be examined on almost identical grounds as
in the case of a single-well one, with an additional
parameter, namely the ratio between  barrier height and chemical
potential, which ranges from a single-well problem for negligibles barriers to
the double degenerate regime where one exactly halves the condensate into two
pieces, in the TF limit of the  GPE for infinite separations.

\acknowledgements
 
We ackowledge fruitful discussions with Dr. Artur Polls. This paper was performed under grants PICT 1706 from Agencia
Nacional de Promoci\'on Cient\'{\i}fica y Tecnol\'ogica 
and PIP 4209/96 from Consejo Nacional de Investigaciones Cient\'{\i}ficas y T\'ecnicas of Argentina.

\section*{Figure Captions}
\begin{description}
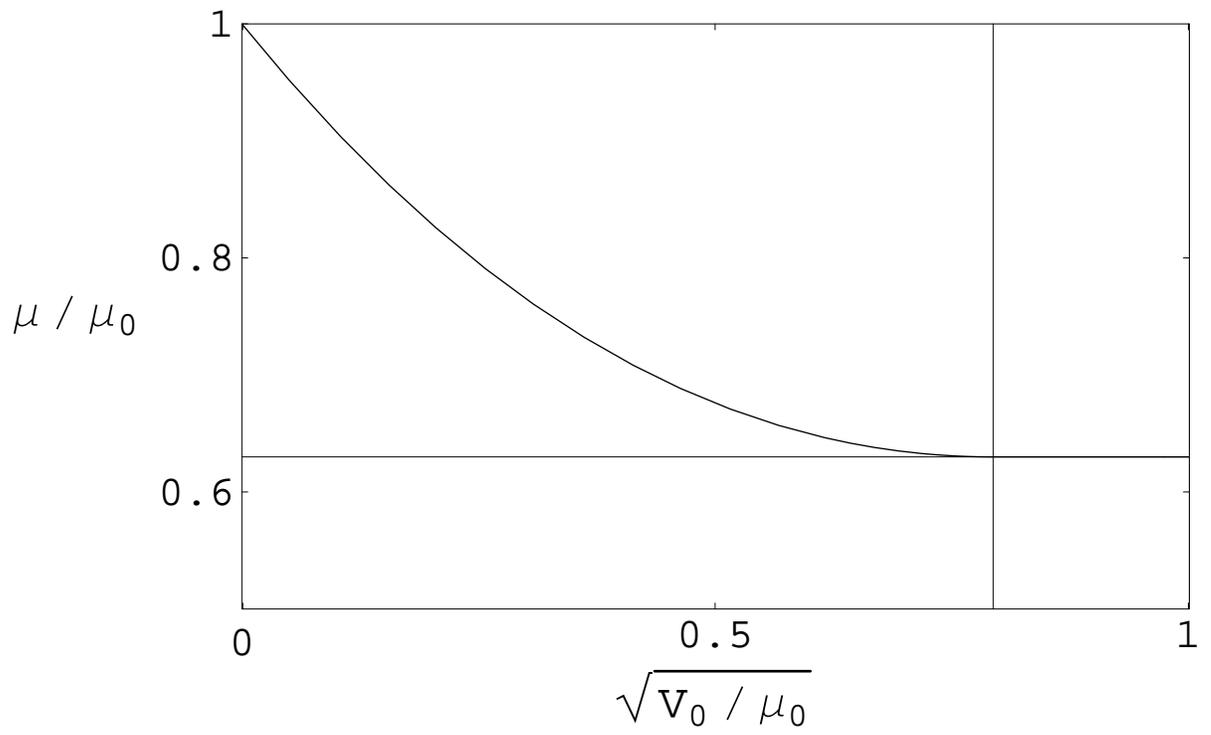
\item[FIG. 1.] The  chemical potential $\mu$ for the 1D harmonic
double well in the TF limit in units of the single well value $\mu_0$ as a 
function of the dimensionless well separation (in units of the size of 
the single-well condensate). The horizontal and vertical thin lines
respectively indicate the chemical potential and barrier height corresponding 
to full splitting into two disjoint single wells.
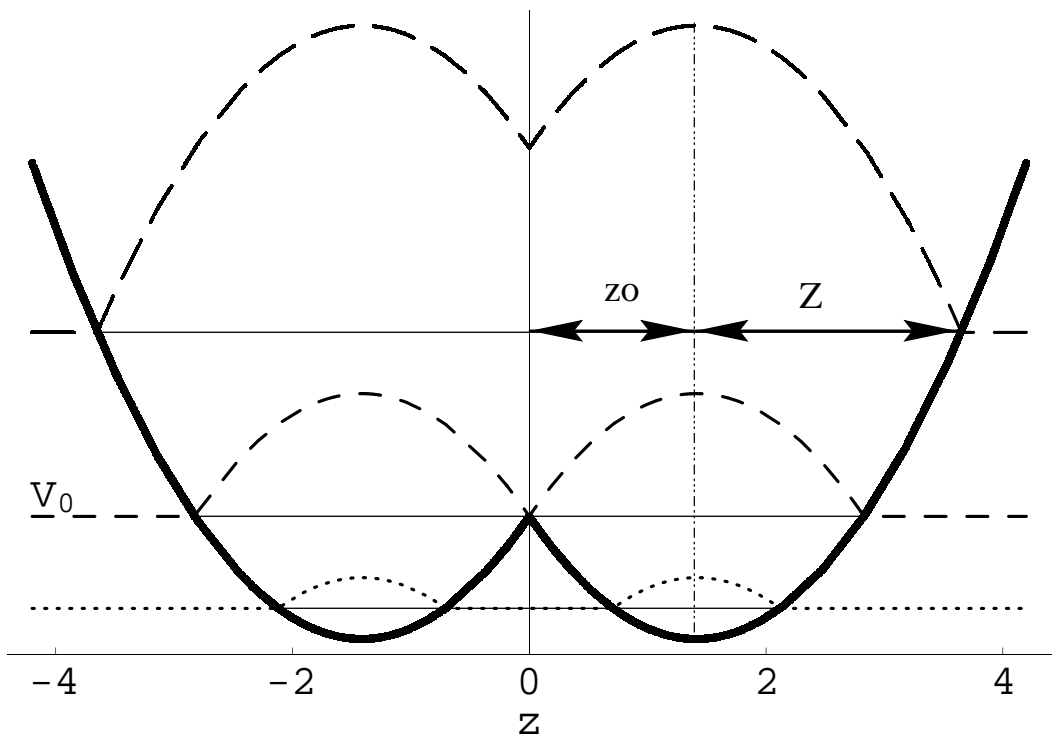
\item[FIG. 2.] Three TF density profiles (in arbitrary units). Full lines indicate 
the value of the chemical potential for which the given contour is a condensate 
edge. The density profiles have been moved up to the corresponding chemical 
potential for easier visualization.
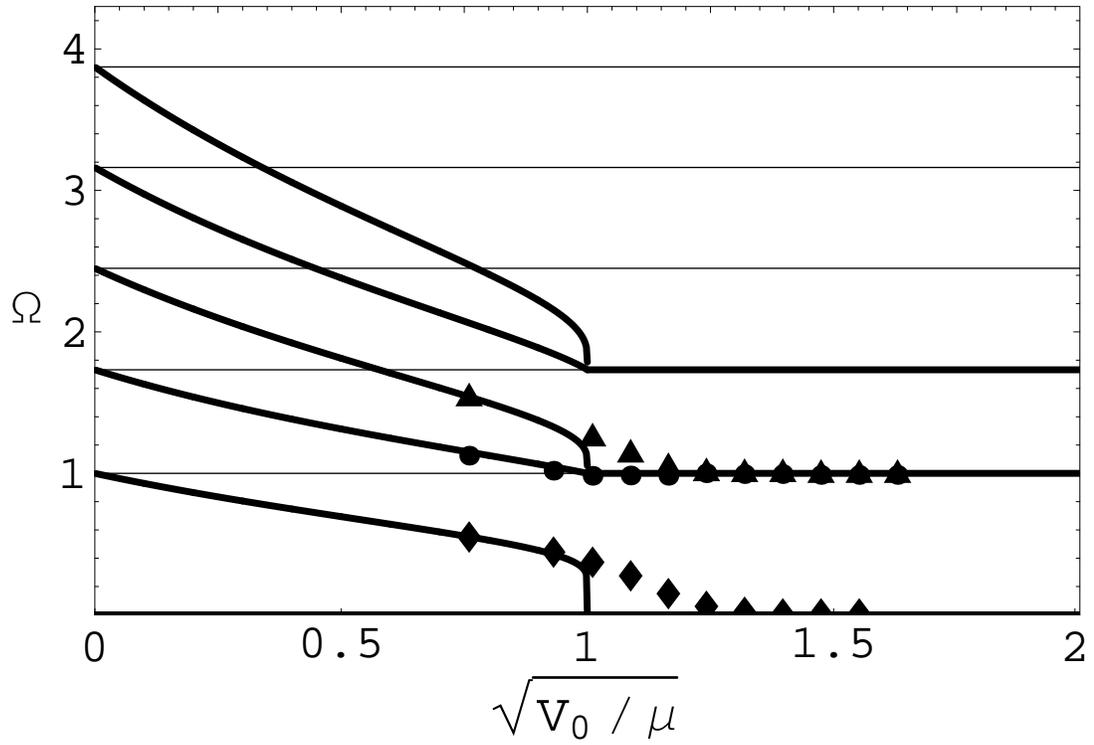
\item[FIG. 3.] Excitation spectrum (in units of $\omega$) of the
1D hydrodynamic equation (\ref{hidro1D}) as a function of the dimensionless
 parameter $\sqrt{V_0/\mu}$. The horizontal thin lines indicate the
spectrum for the single well problem. Some eigenvales provided by the fft 
calculation are displayed with symbols. 
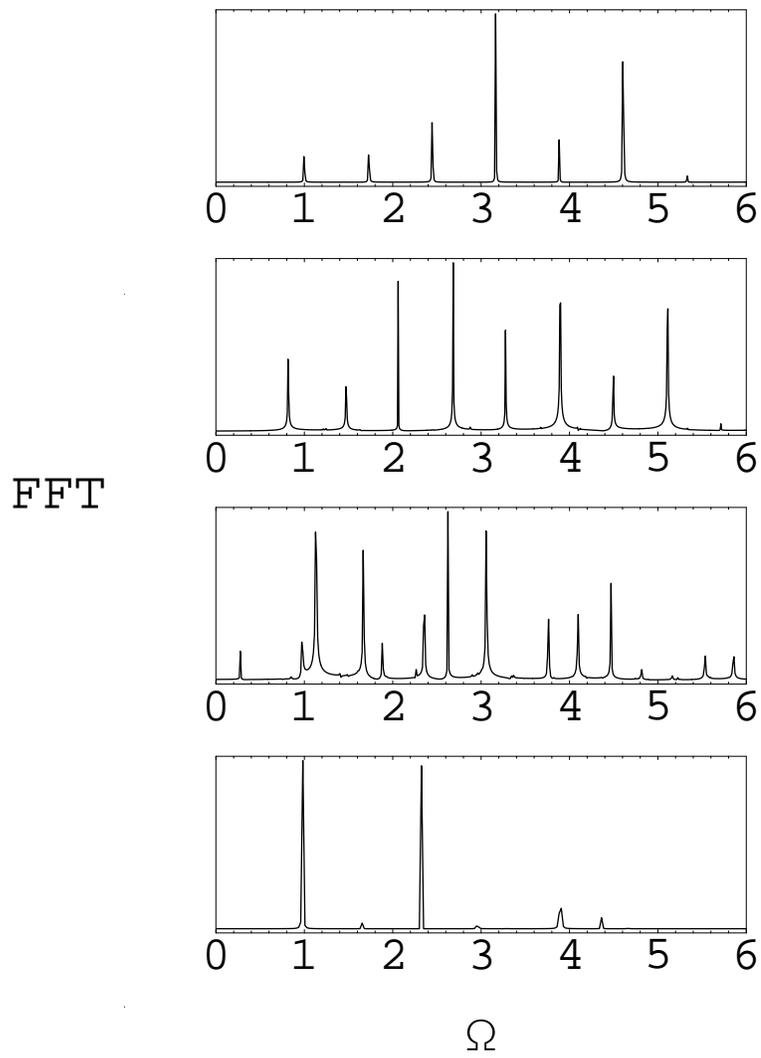
\item[FIG. 4.] The fast Fourier transform of the evolution of a
perturbing density in the 1D-TF limit, for  $\sqrt{V_0/\mu}$= 0, 0.4, 1.09 and  3.9 (from top to bottom). The
intensity of the peaks is displayed in arbitrary units and $\Omega$ is given in units of
$\omega$.
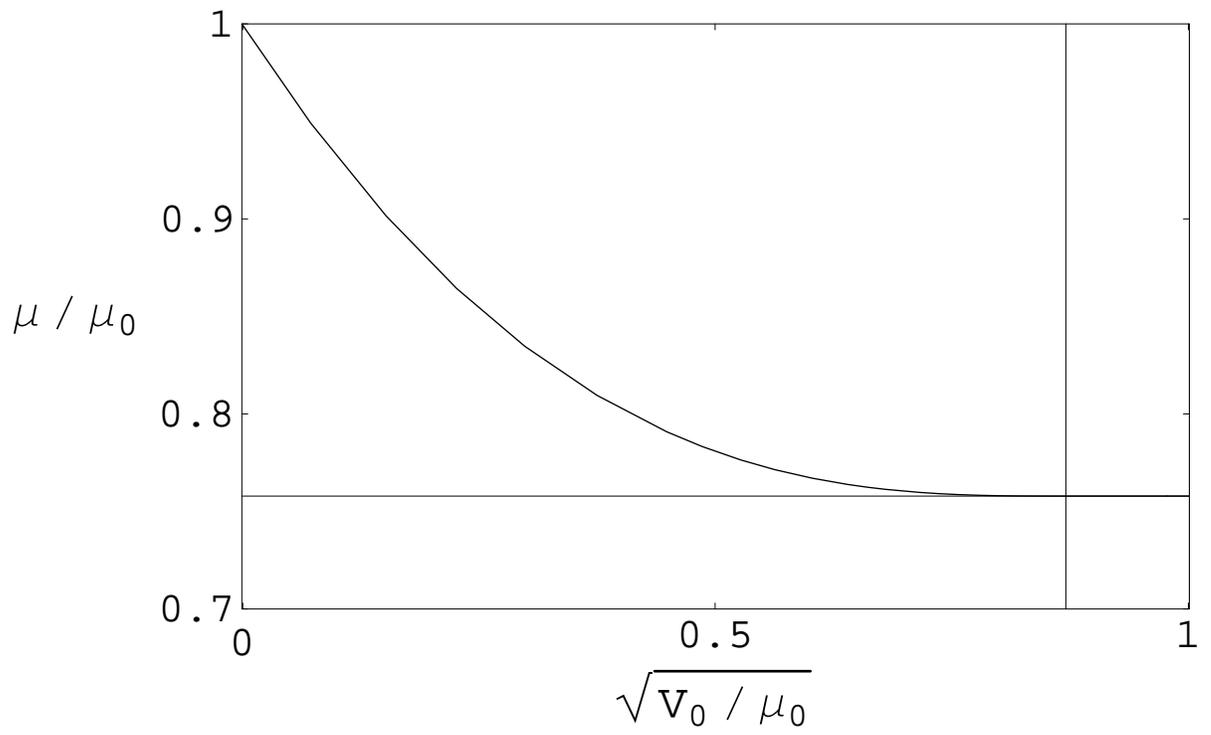
\item[FIG. 5.] Same as Fig. 1 for the 3D double well problem. 
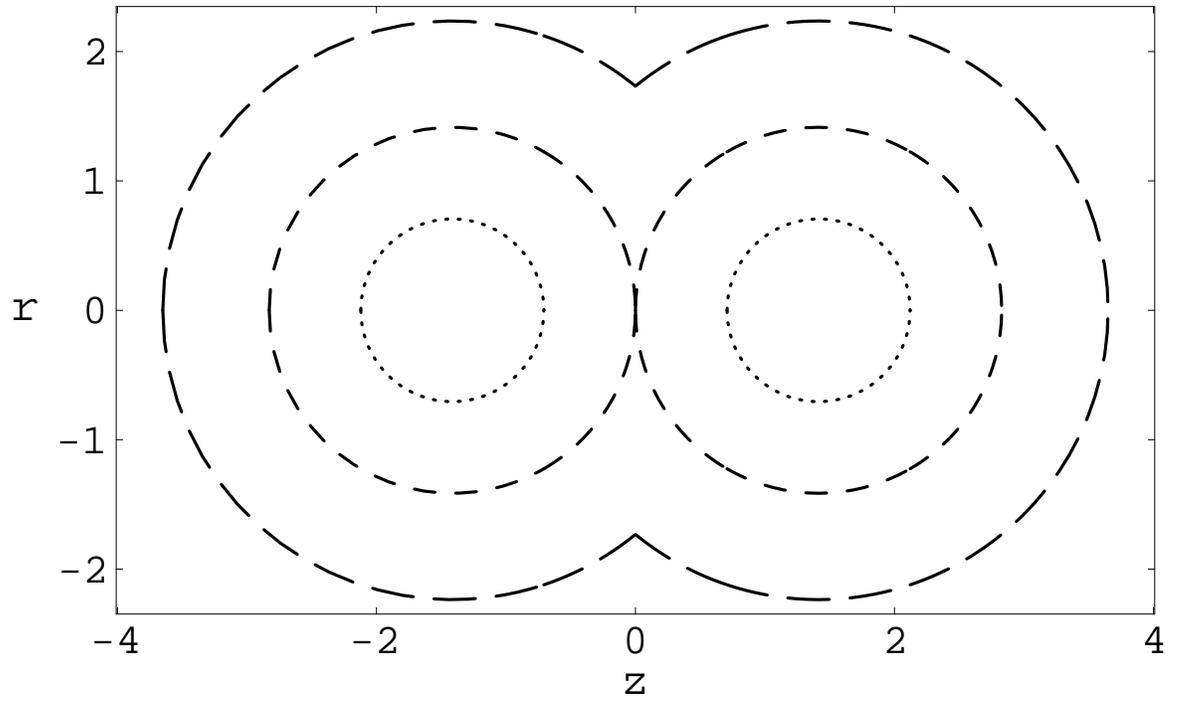
\item[FIG. 6.] Three equiprobability contours (in arbitrary units) 
for a TF condensate, or equivalently, the condensate edges for the same  
chemical potentials considered   in Fig. 2.  
\item[FIG. 7.] Same as Fig. 3 for the
3D hydrodynamic problem: (a)  $m=0$ and
(b) $m=1$. Thin lines correspond to  the single-well spectrum for the same parameters.
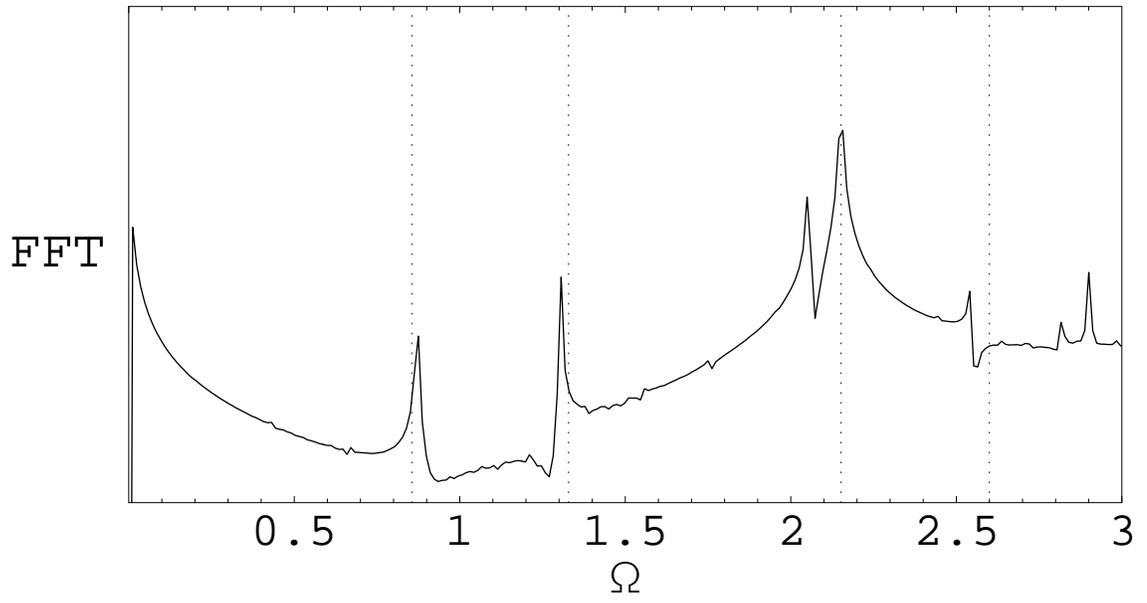
\item[FIG. 8.] Spectrum of the 3D problem provided by the fft
calculation for $\sqrt{V_0/\mu}$=0.14 as a function of $\Omega$ (in units of $\omega$). The vertical lines show a few  hydrodynamic
eigenfrequencies obtained through  diagonalization   of
(\ref{hidro3D}). The height of the peaks is set in arbitrary units.
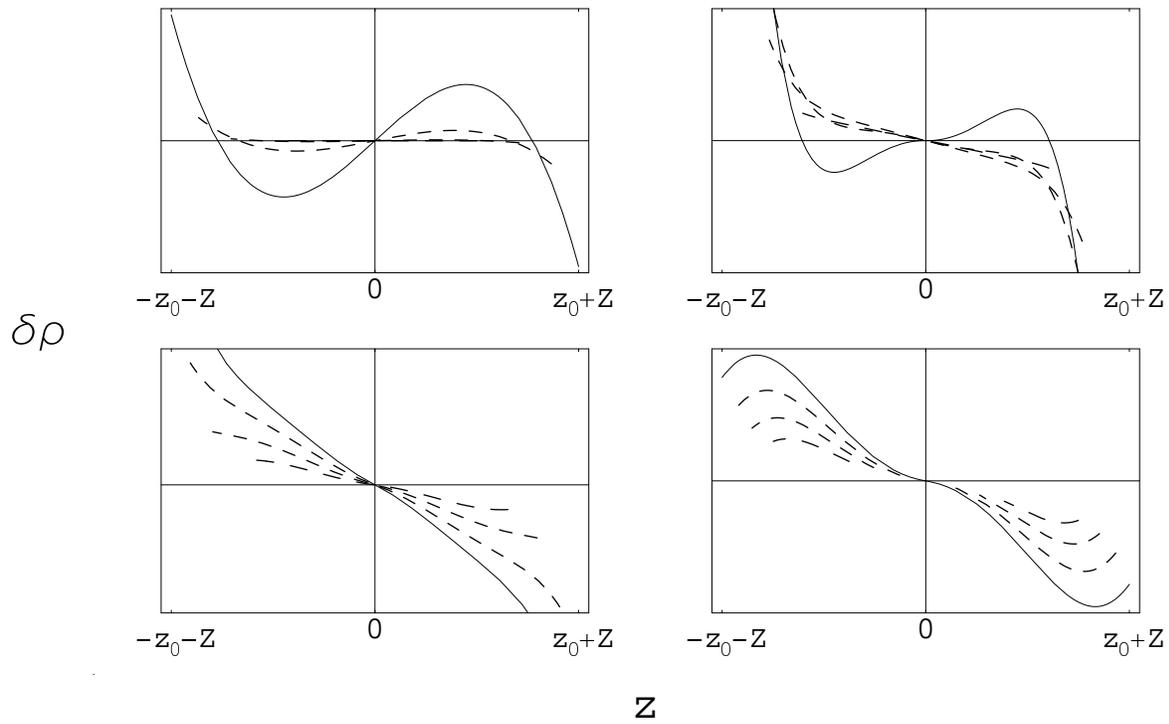
\item[FIG. 9.] The first odd-parity density fluctuation $\delta \rho[r(z),z]$ of the  condensate as a function of $z$ (in arbitrary units), for $m$=0. The four  portraits show the evolution from a single well problem to 
the limiting case $V_0\sim \mu$ with the values  $\sqrt{V_0/\mu}$ =
0.01, 0.26, 0.43 and 0.99. Each curve on each portrait corresponds 
to the contour values $\rho_0/\!\rho_{\rm max}$= 0, 1/4, 1/2, 3/4; the full 
lines denoting the condensate edge and $\pm (z_0+Z)$  points on the z-axis. 
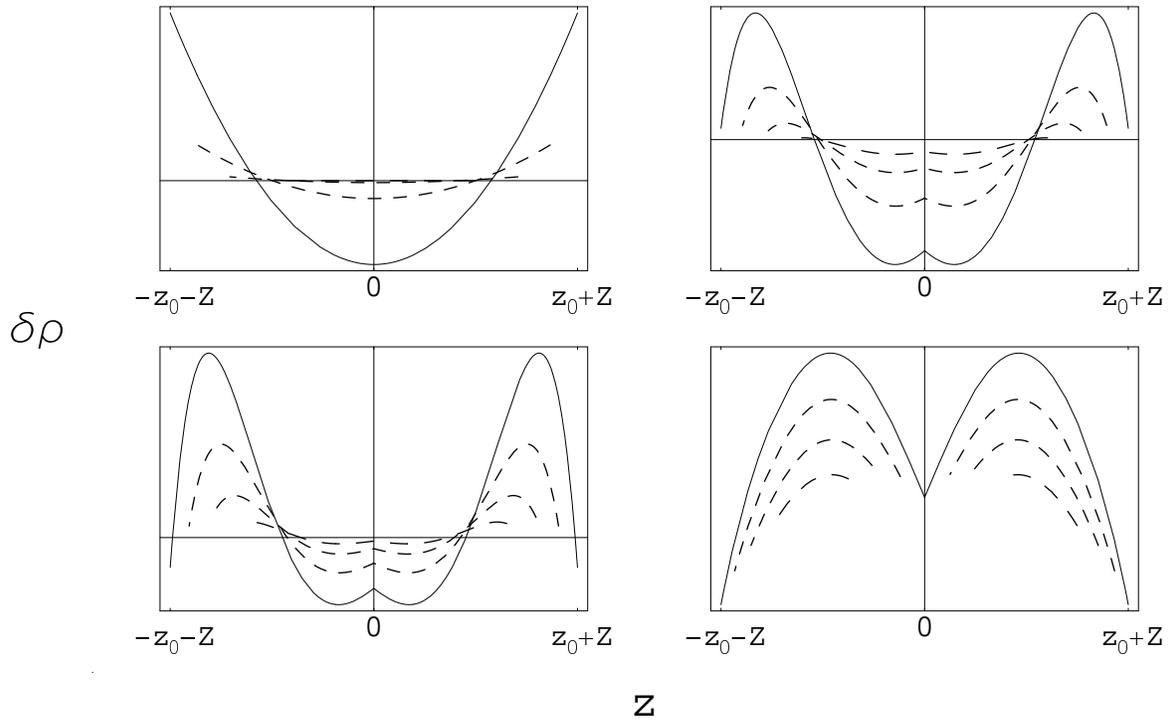
\item[FIG. 10.] Same as Fig. 9 for the first even parity eigenstate.
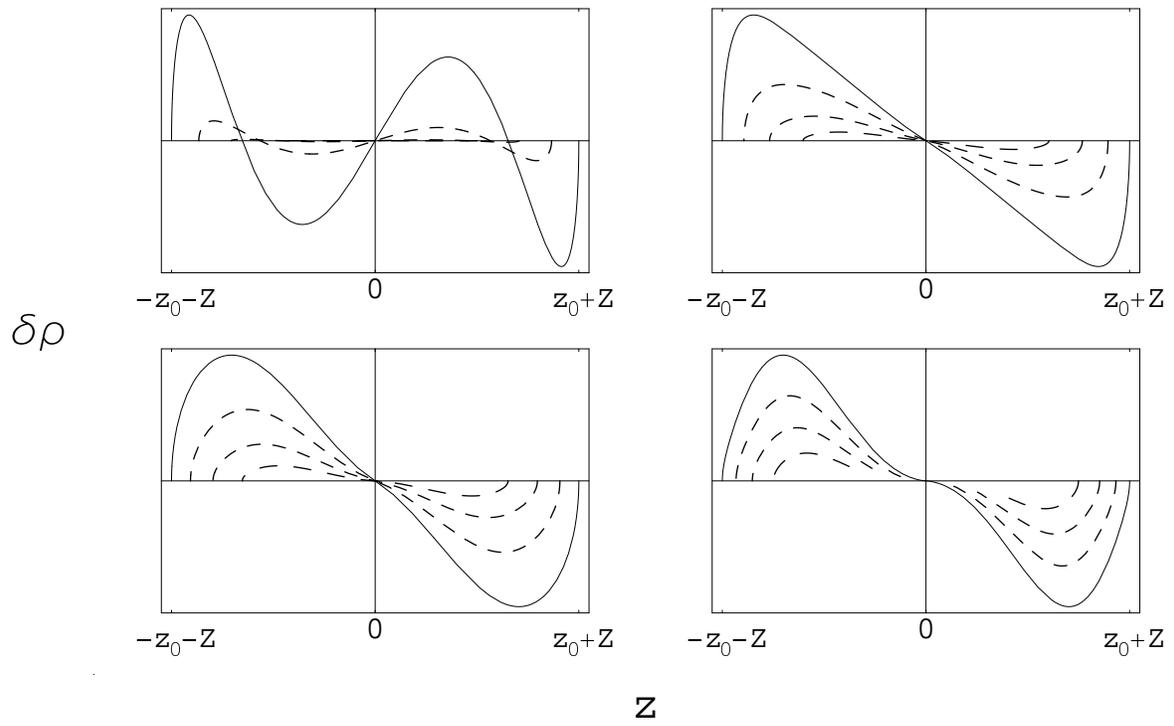
\item[FIG. 11.] Same as Fig. 9 for $m$=1.
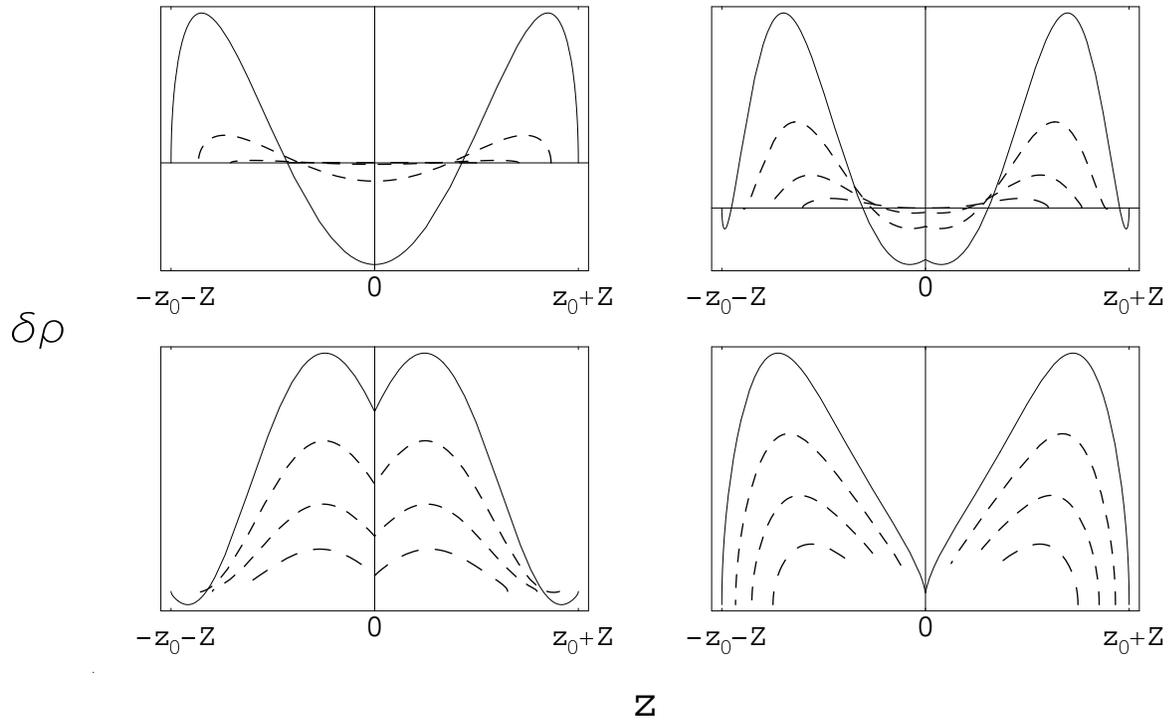
\item[FIG. 12.] Same as Fig. 10 for $m$=1.
\end{description}

\end{document}